\documentstyle[prb,aps,twocolumn]{revtex}
\input{epsf}
\begin{document}

%\draft

\title{Electronic structure, magnetism and superconductivity of MgCNi$_{3}$ }

\author{S.B. Dugdale$^{1,2}$ and T. Jarlborg$^{1}$}

\address{1. D\'epartement de Physique de la Mati\`ere Condens\'ee,
Universit\'e de Gen\`eve, 24 quai Ernest Ansermet, CH-1211 Gen\`eve 4,
Switzerland}

\address{2. H.H. Wills Physics Laboratory, University of Bristol, Tyndall
Avenue, Bristol, BS8 1TL, United Kingdom}

\date{\today}

\maketitle

\begin{abstract}

The electronic structure of the newly discovered superconducting perovskite
MgCNi$_3$ is calculated using the LMTO and KKR methods. The states near the
Fermi energy are found to be dominated by Ni-d. The Stoner factor is low
while the electron-phonon coupling constant is estimated to be about 0.7,
which suggests that the material is a conventional type of superconductor
where T$_C$ is not affected by magnetic interactions. However, the
proximity of the Fermi energy to a large peak in the density of states in
conjunction with the reported non-stoichiometry of the compound, has
consequences for the stability of the results.

\end{abstract}

\pacs{PACS go here}

The perovskite oxides have hosted a wide variety of exotic
phenomena, including high-temperature superconductivity \cite{bednorz:86},
colossal magnetoresistance \cite{helmolt:93} and ferroelectricity
\cite{lines:77}.  However, the recent report of superconductivity below 8K
in the intermetallic compound MgCNi$_{3}$ by He {\it et al.} \cite{he:01}
is the first in a material which has the perovskite structure without any
oxygen. Presented as a three-dimensional analog of the borocarbide
superconductors \cite{cava:94} (where there has recently been suggestion of
non-$s$-wave pairing \cite{yokoya:00}), He {\it et al.} speculated that the
favoring of a superconducting ground state over a ferromagnetic one in
MgCNi$_{3}$ (with its high Ni content), makes it a potential candidate for
unconventional superconductivity.  In the rare-earth nickel borocarbides,
the presence of moment-bearing rare-earth atoms and high Ni content makes
the very existence of superconductivity surprising, and some of the
phenomena associated with its interplay with magnetism have not been
observed in any other superconducting material
\cite{canfield:98,norgaard:00}. Huang {\it et al.} \cite{huang:01},
however, report no magnetic or structural transitions forMgCNi$_{3}$ in the
range 2---295K. Recent tunneling measurements by Mao {\it et al.}
\cite{mao:01} have indicated that MgCNi$_{3}$ is a strong-coupling
superconductor, and moreover that the pairing symmetry could be
non-$s$-wave. Here we present calculations of the electronic structure of
MgCNi$_{3}$, focusing on the central issues of magnetism and
superconductivity.

The structure of MgCNi$_{3}$ has been reported to be that of a classic
perovskite \cite{huang:01}, comprising a C atom at the body-center
position, surrounded by a cage of Ni atoms at the face-center positions,
with the Mg occupying the cube corners. The electronic structure of
MgCNi$_{3}$ was calculated using both the LMTO \cite{andersen:75} and, as a
cross-check and comparison, with the KKR \cite{bruno:97} method. The
lattice parameter was fixed at the experimental low-temperature value of
7.2 a.u. (3.81\AA) \cite{huang:01}. For the LMTO calculations, made within
the atomic sphere approximation (ASA), self-consistency was reached using
286 k-points within the irreducible wedge of the simple cubic Brillouin
zone (BZ).  The KKR code used non-overlapping muffin-tin potentials, and
the mesh of k-points was adaptive and determined principally by the
required integration tolerances \cite{bruno:97}.

\begin{center}
\begin{figure}
\epsfxsize=250pt
\epsffile{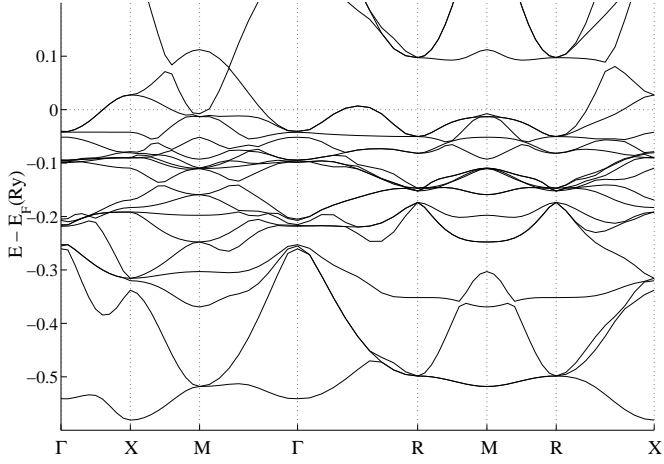}
\caption{Electronic band structure of MgCNi$_{3}$ along the high-symmetry directions
of the simple cubic BZ.}
\label{bands}
\end{figure}
\end{center}

\begin{center}
\begin{figure}
\epsfxsize=250pt
\epsffile{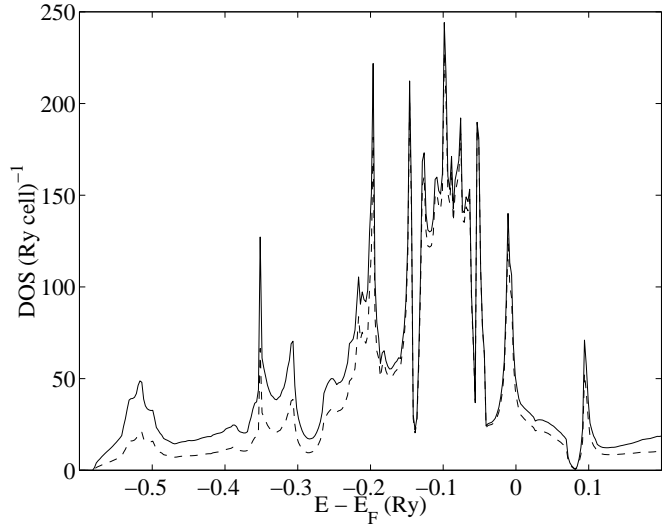}
\caption{Density of states (DOS) for MgCNi$_{3}$. The total DOS (solid
line) and the partial Ni DOS (dashed line) are shown.}
\label{dos}
\end{figure}
\end{center}

\begin{center}
\begin{figure}
\epsfxsize=200pt
\epsffile{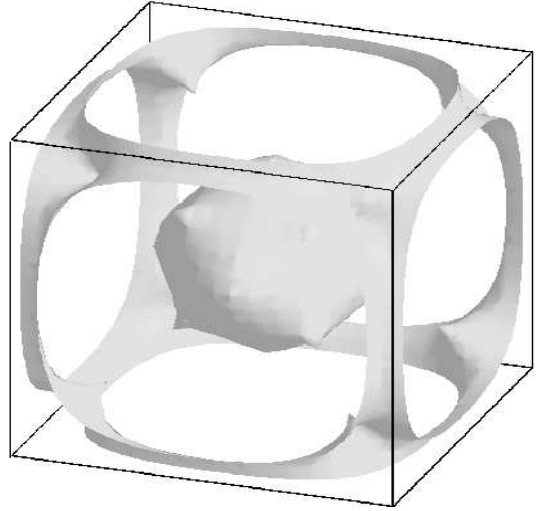}
\epsfxsize=200pt
\epsffile{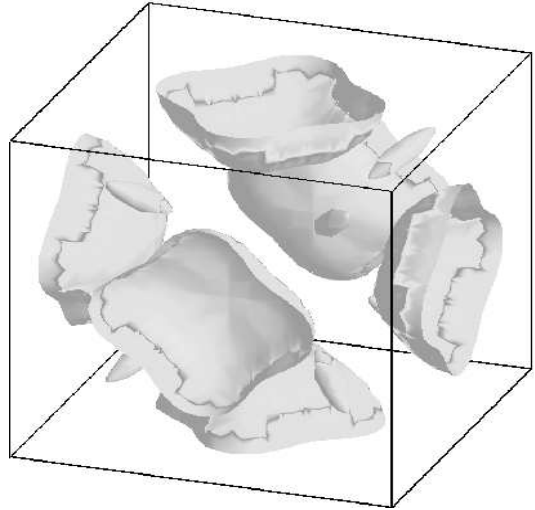}
\caption{The Fermi surface sheets of MgCNi$_{3}$.}
\label{fs}
\end{figure}
\end{center}

The bands, density of states (DOS) and Fermi surface (FS) from the LMTO
results are shown in Figs.~\ref{bands},~\ref{dos} and \ref{fs}
respectively. In Fig.~\ref{dos}, the dashed line represents the
contribution of Ni to the total (solid line) DOS. It should be noted that
near the Fermi energy, $E_{F}$, the DOS is almost completely due to Ni (see
Table~\ref{dosef}).  The KKR results were very similar to those from the
LMTO, but the Ni d-bands were flatter, generating a larger DOS at the Fermi
level (86.4 compared to 47.6 states (Ry cell)$^{-1}$ from the
LMTO). Clearly, with the Fermi level lying in such close proximity to a
large peak in the DOS, there will be a sensitivity in the derived
quantities. Should single crystal samples become available, a determination
of the FS topology would be a stringent check on the location of the
Fermi energy on the DOS peak, since the low dispersion of the Ni-d bands
would make the topology very sensitive to small shifts in $E_{F}$.

Henceforth, we will focus on the LMTO results.  These
calculations include s,p,d and f states for all atoms because the
partial f-DOS is required in the evaluation of the electron-phonon
coupling. Some bands were found to be sensitive to the choice of
linearization energies, but a separate LMTO calculation using only s-,p-
and d-states for Ni, and s and p states for Mg and C, gave the same FS
topology and a very similar total DOS at $E_F$. Two bands cross $E_F$,
making one jungle-gym-like FS sheet around the BZ edge, with a
spheroid-like sheet around $\Gamma$. The second FS plot shows some
X-centered shell-like features at the BZ faces and small, delicate
`cigars' along $\Gamma$-R.

The bands (Fig.~\ref{bands}) are quite dispersive near $E_F$, with a Fermi
velocity of $2.0 \times 10^5$ m/s, which in combination with the large DOS,
would give the material a good metallic conductivity. The temperature
dependence of the resistivity is reported as having the signature of a poor
metal \cite{he:01}. As the calculated DOS is strongly varying near $E_F$,
it is expected that unusual T-dependencies could occur.

The propensity for a metallic system to adopt a ferromagnetic ground state
can be expressed in terms of the Stoner factor, $S=1/(1-\bar{S})$, this
being the exchange enhancement, which diverges at a ferromagnetic
transition \cite{jarlborg:80,gunnarson:76}.  The DOS per Ni atom is
considerably smaller than in fcc Ni, and the calculated value of the Stoner
factor $\bar{S}$=0.43 is far from the ferromagnetic limit. This suggests
that, in contrast to common expectations of Ni-rich compounds \cite{he:01},
magnetism will not be present in this compound and should not interfere
much with superconductivity. Antiferromagnetic ordering would also seem
unlikely, since the three dimensional structure makes the FS topology quite
complicated with no obvious nesting features.

The electron-phonon coupling constant, $\lambda$, can be expressed as 
\cite{dacorogna:84}, 
\begin{equation}
\label{lambda}
\lambda = \sum_{i}\frac{\eta_{i}}{M_{i} \langle \omega_{i}^2 \rangle}.
\end{equation}
where the sum runs over all atoms, $i$, with masses, $M_{i}$, and phonon
frequencies, $\omega_{i}$, while the numerator, $\eta_{i}$, is the
electronic contribution. Here, $\eta$ was calculated in the rigid
muffin-tin approximation \cite{dacorogna:84}. This implies that only dipole
terms without screening are included. The large Ni-d DOS is such that the
contribution to $\eta$ is dominated by Ni d to f scattering.  In order to
get an idea of how large $\lambda$ could be, we estimate the phonon
contribution by using the Debye temperature of fcc Ni (450 K
\cite{kittel:96}). In this case, we obtain 0.67 for $\lambda$, which, with
the bare DOS, gives an electronic specific heat coefficient of 4.7 mJ per
mole Ni K$^{-2}$. This is smaller than the experimental value \cite{he:01}
(about 10 mJ per mole Ni K$^{-2}$), which leaves room for additional
enhancements due to spin- or valence-fluctuations. From their specific heat
measurements, He {\it et al.} \cite{he:01} infer $\lambda=0.77$ (with
errors of +0.17 and -0.09).  We should like to point out that with the bare
Fermi energy DOS from the KKR calculations, the agreement with the
electronic specific heat constant would be much closer. On the other hand,
the LMTO-derived $\lambda$ would give a superconducting critical
temperature, $T_{c}$, of the order of 10K (using the McMillan formula with
$\mu^{*}=0.13$ \cite{mcmillan:68}), in line with the experimental
observations \cite{he:01}, whereas that derived from the KKR DOS would be
much larger. It should also be noted that this calculation of $\lambda$ is
strongly dependent on the phonon frequencies, which are not calculated
here.

Since there is a large peak in the DOS close to $E_F$, and if doping (the
occupancy of the C-site is 0.96 rather than unity, in the measured material
\cite{he:01}) were to add 0.5 holes per unit cell, $E_F$ would reach that
peak. In this case, the DOS at $E_{F}$ would be more than twice as large,
reaching 130 states per cell per Ry, which would mean that $\bar{S}$ would
become close to unity. However, this is unlikely for the measured material,
since the specific heat coefficient would become too large, both due to the
bare DOS and from the additional enhancements.

%It has been pointed out that the essentially filled d-band in Ni is similar
%to the filled O p-bands in perovskite oxides, which may be the stabilizing
%factor for this particular structure \cite{he:01}. However, the band
%structure of oxides like WO$_3$, ruthenates, manganites and titanates with
%the same structure is quite different for this material, and the physical
%properties depend on the presence of a large gap.

The electronic structure of the newly discovered perovskite superconductor,
MgCNi$_{3}$ is reported.  In summary, we find that although the DOS at the
Fermi level is dominated by the Ni d-states, it is not large enough to
induce magnetic instabilities. However, the DOS is sufficiently large to
produce strong electron-phonon coupling.

One of the authors (SBD) would like to thank the Royal Society and the DPMC
(University of Geneva) for financial support.

\begin{table}
\caption{Site decomposition of the density of states at $E_{F}$ [(Ry
cell)$^{-1}$].}

\label{dosef}
\begin{tabular}{ccccc}
Site & s & p & d & f \\
\tableline
Mg        & 0.1 & 2.0 & 0.5  & 0.1 \\
C         & 0.2 & 3.6 & 0.1  & 0.2 \\
Ni$_{3}$  & 1.0 & 2.8 & 36.3 & 0.3 \\
\end{tabular}
\end{table}

\end{document}